# Benchmarking Machine-Learning Interatomic Potentials for Dynamical Stability in Inorganic Semiconductor Nanocrystals: A CdSe Case Study

**Muhammed Usman[1], Masuma Suleymanova[1], Zain Ul Abideen[1], Mario Fernández-Pendás[2*], and Ivan Infante[1,3,*]**

[1] BCMaterials, Basque Center for Materials, Applications, and Nanostructures, UPV/EHU Science Park, Leioa, 48940 Spain
[2] Departamento de Metodos Cuantitativos, Facultad de Economia y Empresa, University of the Basque Country (UPV/EHU), Leioa, Spain
[3] Ikerbasque Basque Foundation for Science, Plaza Euskadi 5, Bilbao, 48009 Spain

*Author to whom any correspondence should be addressed.

**E-mail:** mario.fernandez@ehu.eus ; ivanatoniocarlo.infante@ehu.es ; ivan.infante@bcmaterials.net

## Abstract

Machine-learning interatomic potentials (MLIPs) enable nanosecond-scale atomistic simulations of inorganic semiconductor nanocrystals, but low errors on held-out configurations do not necessarily guarantee stable molecular dynamics. We benchmark five graph-neural-network MLIPs, SchNet, PaiNN, NequIP, Allegro and MACE, for dynamical stability in a chloride-passivated cadmium selenide nanocluster containing 149 atoms. The models were trained under harmonized conditions on 1,000 configurations and evaluated against 2,000 held-out configurations generated using density functional theory, considering force accuracy, computational efficiency, uncertainty and structural stability during 1 ns simulations at 300 K. Allegro produced the lowest validation force mean absolute errors, ranging from 12 to 14 meV Å$^{-1}$, whereas SchNet produced the largest, ranging from 84 to 110 meV Å$^{-1}$. This ranking did not predict dynamical robustness: NequIP remained stable for 1 ns without additional training data, whereas MACE became stable only after an ensemble-based active-learning procedure added 34 uncertainty-selected configurations. PaiNN and Allegro remained unstable after the addition of 100 and 95 configurations, respectively, and SchNet also failed to achieve stable dynamics within the tested augmentation budget. Under the benchmark hardware conditions, the machine-learning potentials required 6–65 ms per molecular-dynamics step, compared with approximately 20 s for density functional theory. These results show that held-out prediction errors alone are insufficient for selecting interatomic potentials for finite, surface-dominated nanostructures. Reliable deployment requires long-timescale dynamical testing combined with uncertainty-guided refinement, while the effectiveness and data efficiency of active learning remain strongly architecture dependent.

## 1. Introduction

Quantum dots (QDs) are colloidal semiconductor nanocrystals that display size-dependent optical and electrical characteristics due to quantum confinement[1,2]. Their tunable bandgaps, discrete electronic states, and photoluminescence quantum yields (PLQYs) approaching unity have established QDs as versatile materials for optoelectronic applications. They are now employed or actively investigated in technologies including commercial displays[3] high-efficiency photovoltaics, broadband photodetectors[4], and single-photon sources for quantum photonics[5]. Despite their considerable technological potential, the performance of QDs is often limited by surface defects[6]. Their high surface-to-volume ratio places a large fraction of atoms at or near the surface, where incomplete passivation, dangling bonds and local structural distortions can introduce localized electronic states within or near the bandgap[7–11]. These states may trap photogenerated electrons or holes, thereby promoting non-radiative recombination and reducing PLQY and carrier-transport efficiency[12]. Establishing direct relationships between these defects and their electronic signatures remains challenging because QD surfaces are structurally heterogeneous and dynamic[13], while conventional ensemble-averaged characterization techniques generally lack the combined spatial, temporal, and chemical resolution required to resolve individual trapping sites[1,14].

Computational modelling provides a complementary route to resolving the atomistic origin and dynamics of surface defects[7,15,16]. Achieving this goal, however, requires an accurate description of chemically complex QDs and their structural evolution. Elemental analyses commonly indicate that QD inorganic cores can be non-stoichiometric and cation-rich, with charge neutrality provided by anionic ligands and colloidal stability maintained by the surrounding ligand shell[17–20]. Their heterogeneous surfaces create diverse local chemical environments involving undercoordinated atoms, mobile ligands, and potentially labile bonds. Density functional theory (DFT) is the standard approach for obtaining a quantum-mechanical description of such systems[21,22] but conventional Kohn–Sham DFT exhibits steep cubic scaling with the number of basis functions[23]. Experimentally relevant colloidal QDs contain hundreds to thousands of atoms, requiring tens of thousands of basis functions. Consequently, extensive structural relaxation and molecular dynamics (MD) simulations over experimentally meaningful timescales remain computationally prohibitive. Empirical classical force fields offer substantially lower, often linear, scaling and can readily access larger systems and longer timescales[13,24–26]. However, they generally lack the chemical flexibility and reactive transferability needed to describe bond rearrangements, ligand mobility, and the diverse coordination environments characteristic of QD surfaces.

Recent advances in machine-learning interatomic potentials (MLIPs) offer a promising route to bridging the accuracy–efficiency gap between first principles and empirical methods[27–30]. By learning the relationship between atomic structure and the quantum-mechanical potential-energy surface, MLIPs can approach DFT-level accuracy at a substantially lower computational cost. Early models relied on fixed, manually designed descriptors of local atomic environments and often required extensive system-specific tuning[31]. More recent approaches employ graph neural networks (GNN) that learn structural representations directly from atomic neighbourhoods through message passing[28–30]. Equivariant architectures explicitly enforce the invariance of predicted energies and the corresponding rotational covariance of atomic forces, enabling a more expressive description of complex and dynamically evolving atomic environments[30–32]. These developments make MLIPs especially attractive for modelling chemically heterogeneous QD surfaces over length and time scales that remain inaccessible to direct DFT simulations.

Despite the rapid expansion of high-throughput DFT datasets and GNN architectures, standardized frameworks for benchmarking MLIPs remain limited[33]. Recent years have also seen the emergence of broadly transferable or “foundation” machine-learning potentials, pretrained on large databases spanning diverse chemical and configurational spaces[34–36]. However, their training and evaluation have focused primarily on periodic materials and relatively idealized surfaces or slabs[37,38]. Their performance for finite, ligand-passivated semiconductor nanocrystals remain comparatively unexplored.

In this context, colloidal QDs present a demanding test of MLIP transferability. Their high surface-to-volume ratios create strongly distorted coordination environments, while intersections between crystallographic facets generate undercoordinated edges and vertices. Their surfaces may also undergo bond rearrangements and ligand migration as they evolve structurally. Such structural features and fluctuations are directly relevant to trap formation, which is frequently associated with recognizable geometric fingerprints, including highly undercoordinated atoms, displaced surface species and locally strained bonding environments. Although conventional MLIPs do not explicitly predict electronic states, they can track the formation, evolution and persistence of these trap-associated motifs over timescales inaccessible to direct DFT simulations. QDs therefore provide a stringent benchmark for assessing whether MLIPs can reproduce complex surface dynamics involving structural environments that are poorly represented in current universal datasets[33] .

Here, we systematically benchmark five widely used GNN-based interatomic potentials, SchNet, PaiNN, MACE, NequIP and Allegro, for CdSe nanocrystals[27,29,30,32,39]. CdSe is a prototypical semiconductor QD whose structure, photophysics and surface chemistry have been extensively investigated experimentally and theoretically[40–43]. To enable a controlled comparison, we train, validate, and test all architectures using the same custom DFT dataset, data partitions and evaluation criteria. We evaluate their energy and force accuracy, computational efficiency and stability in molecular dynamics simulations. We further examine whether uncertainty-driven active learning can identify underrepresented configurations and improve predictive accuracy and dynamical stability. This benchmark clarifies the strengths and limitations of invariant and equivariant GNN architectures for finite semiconductor nanocrystals and establishes a reproducible protocol for evaluating MLIPs on chemically heterogeneous QD surfaces[33].

## 2. Methodology

The benchmarking framework was designed to ensure strict consistency across all models at every stage: data preprocessing, training, and post-processing (i.e. MD simulations).

### 2.1. Generation of the Reference CdSe Dataset.

The reference dataset was constructed using a discrete CdSe QD model approximately 2 nm in diameter (**Figure 1**, **Panel 1**). This size was selected to capture the characteristic physical and electronic properties of CdSe quantum dots while remaining computationally tractable for high-level quantum chemical calculations. Although this model sits at the lower bound of typical experimental synthetic sizes (usually 3-4 nm), it is sufficiently large to exhibit a realistic and representative surface. The cluster comprises 149 atoms in total, featuring a non-stoichiometric core of 68 Cd and 55 Se atoms. Assuming formal oxidation states of +2 for Cd and -2 for Se, the resulting excess positive charge was compensated via passivation with 26 chloride ligands (-1 charge) to stabilize surface dangling bonds resulting in the neutral stoichiometry $Cd_{68}Se_{55}Cl_{26}$. Structurally, the nanocrystal exhibits a truncated cuboctahedral morphology, defined by a specific spatial arrangement where six Cd-rich {100} facets, four cation-rich {111} facets, and four anion-rich {-1-1-1} facets intersect. The $Cl^-$ ligands completely passivate the {100} facets that restores the undercoordinated, di-coordinated surface Cd ions to a bulk-like coordination environment and partially passivate the cation-rich {111} facets. As established in previous literature, this specific passivation strategy yields a stable, wide bandgap free of mid-gap trap states, making this system a useful, controlled prototype for benchmarking atomistic potentials[7,24].

To explore the potential energy surface and generate the reference dataset, Born-Oppenheimer Ab Initio Molecular Dynamics (AIMD) simulations[44] were performed in the NVT ensemble at 300 K, with the temperature regulated by a Canonical Sampling through Velocity Rescaling (CSVR) thermostat (time constant of 250 fs)[45]. Using an integration timestep of 2.5 fs, the system was initially equilibrated for 2.5 ps (1000 steps) to ensure thermal stability. Following equilibration, a 7.5 ps (3000 steps) production run was performed. Total energies, atomic forces, and velocities were recorded at every MD step during the production phase, yielding a pool of 3,000 distinct frames that comprehensively capture realistic thermal fluctuations and structural rearrangements. All quantum mechanical reference data were calculated using DFT within the Gaussian Plane Wave (GPW) formalism, as implemented in the CP2K 2024.1 package[46] non-periodic boundary conditions were explicitly enforced for the isolated nanocrystal using a 32x32x32 $Å^3$ simulation cell paired with a Multipole Poisson solver. Electronic exchange-correlation effects were treated with the HLE17 meta-GGA functional[47], selected for its specialized accuracy in describing the structural and band gaps of semiconducting nanostructures. Core electrons were represented using Goedecker-Teter-Hutter (GTH) pseudopotentials[48–50], while the valence space was expanded via optimized molecular double-zeta valence polarized (DZVP-MOLOPT) basis sets[51], resolved with a plane-wave density cutoff of 400 Ry. To construct a diverse training set while limiting the computational cost of model optimization, the 3,000 configurations were first screened for anomalous structures using an isolation-forest algorithm. Their atomic environments were then mapped into a high-dimensional descriptor space and analysed using principal component analysis and k-means clustering. We selected 1,000 representative configurations from the resulting clusters for model training, thereby reducing redundancy while retaining the principal variations in bond lengths, coordination environments and surface structures. The remaining 2,000 configurations were retained as a held-out validation set and were excluded from both parameter fitting and active-learning acquisition. This procedure ensured that the training data eliminated redundancy and maximized the diversity of bond lengths and surface reconstructions across the $Cd_{68}Se_{55}Cl_{26}$ system.

### 2.2 Model Training Setup

The identical dataset of chloride-passivated CdSe nanocluster configurations having 1000 frames was used to train five machine-learning interatomic potentials, spanning both rotationally invariant and equivariant architectures: SchNet, PaiNN, MACE, NequIP, and Allegro. Throughout training, a uniform effective batch size of 8 and a local cutoff radius of 12Å were maintained across all models to ensure a consistent comparison. The structural frameworks of these MLIPs handle spatial directional information differently based on their underlying symmetry constraints. As a purely rotationally invariant model, SchNet relies entirely on scalar atomic environments, mapping interatomic distances via 40 radial basis functions (RBFs) without explicitly updating directional vectors during the message-passing phases. In contrast, PaiNN, NequIP, Allegro and MACE use rotationally equivariant representations that preserve directional information under rotations. Within this group, PaiNN propagates equivariant directional vectors, whereas the higher order equivariant frameworks of NequIP, Allegro, and MACE employ a spherical harmonics angular momentum restriction of $l_{\max} = 1$. To define these local environments, NequIP and Allegro utilize 10 Bessel basis functions coupled

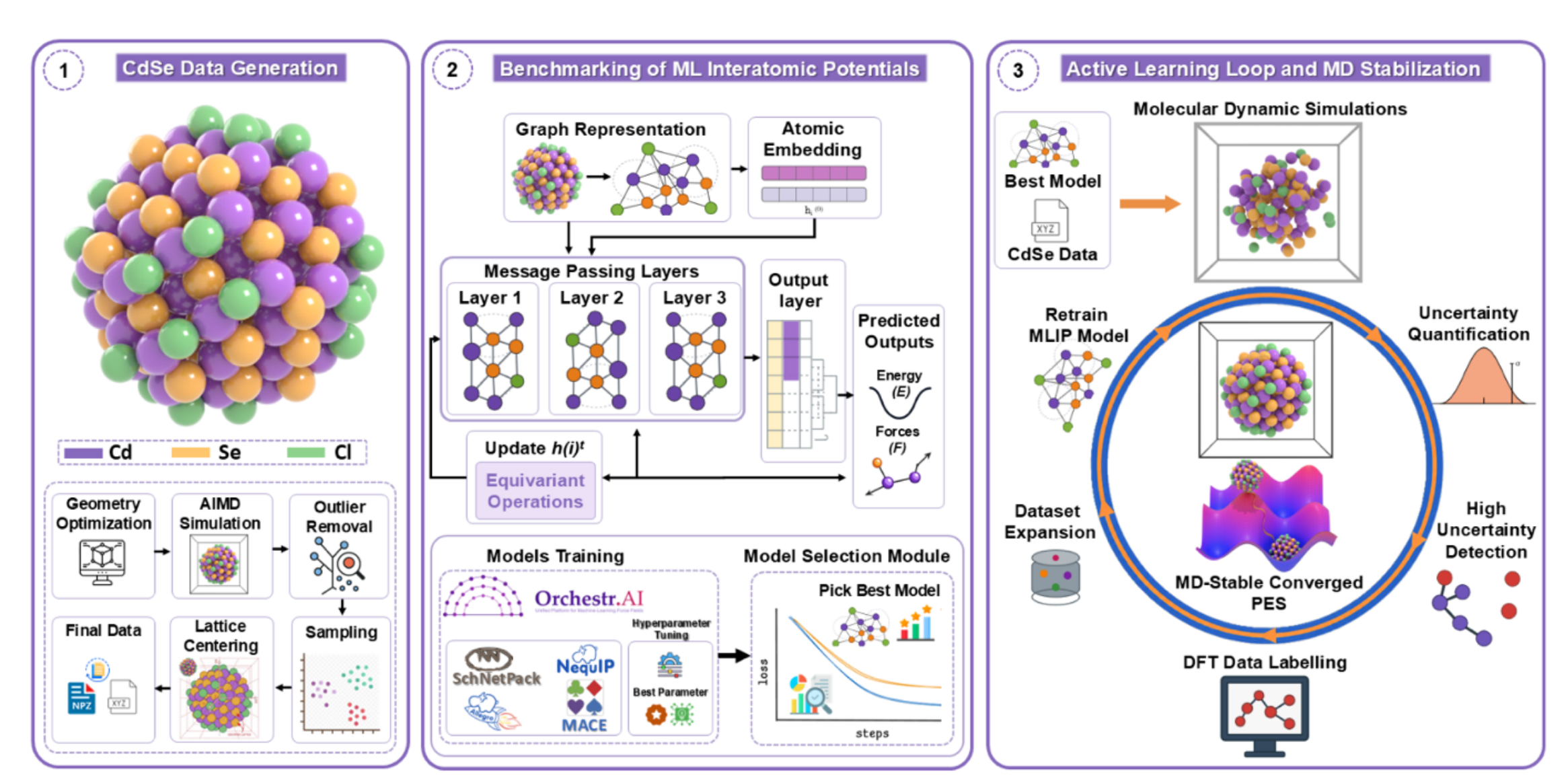

**Figure 1.** Implementation of the Orchestr.AI computational workflow for the development of dynamically stable MLIPs. The successful execution of the pipeline is divided into three key phases presented in this section: **(1)** The generation and statistical clustering of the 149-atom CdSe-Cl reference dataset to ensure a highly diverse structural phase space. **(2)** The standardized benchmarking results evaluating the trade-offs between invariant and equivariant message-passing architectures (SchNet, PaiNN, MACE, NequIP, Allegro). **(3)** The outcomes of the active-learning loop, demonstrating how autonomous uncertainty quantification successfully stabilized the molecular dynamics trajectories to achieve long-term dynamical stability

with a sixth-order polynomial cutoff, whereas MACE uses 40 radial basis functions. Accounting for these equivariant tensor features is critical for capturing the complex bonding environments dictated by the widely varying coordination numbers of the inorganic ions (Cd and Se) between the core and the surface. Furthermore, these representations can encode the directional geometries associated with Cd–Se and Cd–Cl interactions and heterogeneous surface coordination environments. This structural capability ensures that both primary covalent and highly ionic bonds, alongside secondary coordination effects at the intersecting facets, are represented with high physical fidelity.

To prioritize force precision, which is crucial for robust MD simulations, all models minimized a combined loss function dominated by atomic forces using a force-to-energy weighting ratio of 0.95**:**0.05. While MACE, NequIP, and Allegro were trained at a learning rate of $1 \times 10^{-3}$, which provided faster convergence without sacrificing precision, SchNet and PaiNN employed a learning rate of $1 \times 10^{-4}$ for training stability. MACE, NequIP, and Allegro were trained with the Adam optimizer, whereas SchNet and PaiNN were trained with the AdamW optimizer[55,56]. All architectures were able to reach convergence after 250 epochs of training for each model.

**2.3 Benchmarking and Evaluation.**

We assessed every trained model using the DFT reference dataset in terms of training wall time, convergence behaviour, and mean absolute error (MAE) for both forces and energy. The atomic embedding dimension and network depth were varied from 64 to 256 features and three to five layers, respectively, to assess the impact of model size on performance. The trade-off between computing efficiency and precision was measured by this methodical sweep. To guarantee complete reproducibility, all models were benchmarked under identical hardware conditions on our recently developed Orchestr.AI platform[57], utilizing NVIDIA A100 GPUs and 32-core CPU nodes. The Orchestr.AI automated pipeline systematically captured validation losses, training curves and GPU utilization metrics for each architecture**.**

To evaluate the quality of the trained MLIP, we performed MD analysis by selecting one model for each architecture with similar settings: atomic embedding size of 128 and four interaction layers to balance accuracy and computational efficiency (*vide infra*). Simulations were conducted in the NVT ensemble at 300 K using a Langevin thermostat (friction coefficient of 0.01 $fs^{-1}$) and a 1 fs time step for a total of 1,000,000 steps (1 nanosecond). These trajectories enabled an assessment of long-time structural stability, the absence of runaway energies or unphysical distortions, and the evolution of atomic arrangements under the tested thermostat and timestep. Within this benchmark, a trajectory was classified as dynamically stable when it completed the full 1 ns simulation without numerical failure or activation of the predefined geometry and force-cap checks (*vide infra*). Stability therefore refers specifically to the tested initial structure, temperature, thermostat, integration timestep and simulation duration.

To iteratively enhance model robustness and expand structural coverage beyond the initial training pool, a batch-based active learning (AL) workflow was deployed. For each architecture, an ensemble of 10 independent models was trained with randomized data splits to establish a baseline for evaluating epistemic uncertainty across the potential energy surface. During the 1 ns exploratory molecular dynamics trajectories, the prediction epistemic uncertainty $\sigma$ was continuously monitored and computed as the standard deviation of the atomic force predictions across the ensemble members, and an isotonic regression model was applied to calibrate the raw ensemble uncertainty against the observed prediction errors. Acceptance criteria for each architecture were derived from the training-set uncertainty distribution, subject to minimum values of 1.0 meV/atom for the per-atom energy uncertainty ($\sigma E$/atom) and 0.150 eV/Å for the maximum atomic force uncertainty ($\sigma F_{max}$), together with a physical force ceiling of 1.5 times the largest force magnitude in the training set. For PaiNN, the maximum force criterion was set to 0.100 eV/Å, corresponding to approximately the 98th percentile of its force-uncertainty distribution over the exploratory trajectory. Frames exceeding any of these criteria, typically corresponding to highly distorted out-of-equilibrium configurations, undercoordinated edge structures or ligand dissociation pathways, were automatically flagged, and the final acquisition batch was drawn from the flagged pool by a diversity-based ranking to avoid redundant configurations. These structures were then subjected to high-throughput single-point DFT reference calculations with the orbital-transformation method and the same electronic-structure settings employed to generate the original dataset. This DFT labelling yielded accurate total energies and atomic forces, after which the resulting data were appended to the training set and the 10-model ensemble was retrained. This cycle was repeated until either the force prediction uncertainty across trajectories and the global validation errors stabilised, or a predefined budget of active-learning iterations was exhausted, allowing the data efficiency of each architecture to be assessed by the amount of augmentation required to reach stability.

#### 2.4 Computational Framework.

All model training, dataset preprocessing, and subsequent atomistic simulations were standardized using Orchestr.AI, a unified Python-based automation pipeline developed in-house. This centralized framework orchestrates the entire simulation lifecycle under an identical configuration scheme, enforcing a common workflow and configuration scheme across the five codebases: SchNet, PaiNN, NequIP, Allegro and MACE backends. Within this execution environment, the trained MLIP were interfaced with the Atomic Simulation Environment (ASE)[58], which served as the primary driver for molecular dynamics trajectories. Supporting scientific libraries, including NumPy, Pandas, and Scikit-learn, were leveraged within the pipeline to handle robust numerical analysis and automated data extraction. To balance computational throughput and memory efficiency, all MLIP training and benchmarking calculations were executed on NVIDIA A100 GPUs utilizing mixed-precision (float32) arithmetic. Computations were distributed across two high-performance computing (HPC) facilities: the Hyperion cluster at the Donostia International Physics Center (DIPC, Spain) and the Leonardo Booster supercomputer at CINECA (Italy). GPU-accelerated nodes with highly optimized CUDA and PyTorch backends ensured consistent execution speeds and rigorous reproducibility across all tested architectures. The entire automated workflow, spanning initial dataset parsing, pre-model configuration, active-learning logging, and post-processing was controlled via modular YAML-based input files, guaranteeing transparent and repeatable execution. Deployment via the Orchestr.AI framework provides a robust, standardized methodology for training, validating, and smoothly integrating highly transferable MLIPs for complex semiconductor nanomaterials.

## 3. Results and Discussion

The complete computational workflow, alongside a high-resolution atomistic rendering of the 149-atom $Cd_{68}Se_{55}Cl_{26}$ QD model evaluated in this study, is illustrated in **Figure 1**. Following this automated pipeline, the results demonstrate how different GNN architectures balance accuracy, speed, and stability for this complex nanoscale system.

#### 3.1 Benchmarking of Graph Neural Network Potentials for $Cd_{68}Se_{55}Cl_{26}$ QD

As anticipated in **Section 2.3**, to rigorously benchmark the selected GNN interatomic potentials, SchNet, PaiNN, MACE, NequIP, and Allegro were trained on an identical dataset of chloride-passivated CdSe configurations. By employing this standardized protocol, we ensured that any observed variations in performance resulted from inherent architectural differences rather than inconsistent data splitting or training hyperparameter settings. These findings (**Figure 2** and **Table 1**) demonstrated clear trends in accuracy and efficiency. Because inherent DFT errors typically exceed the sub-meV scale, we report our force errors in meV $Å^{-1}$. Across the entire hyperparameter space, Allegro achieved the lowest validation force MAEs, ranging consistently from 12 to 14 meV $Å^{-1}$, while achieving stable training under the tested settings. NequIP and MACE achieved comparable precision, with MAEs between 19 and 31 meV $Å^{-1}$. Notably, despite the complex tensorial operations required for higher order equivariant message passing, the training times for MACE remained computationally manageable thanks to the integration with highly optimized cuEquivariance backends[27]. PaiNN provided a balanced mid-tier performance, achieving MAEs between 34 and 52 meV $Å^{-1}$ at a modest computational cost. SchNet exhibited the largest force errors, ranging from 84 to 110 meV $Å^{-1}$, under the present dataset and training protocol, suggesting that its distance-based invariant representation was less effective for this heterogeneous surface environment.

To thoroughly evaluate the network capacity across the backends, we systematically moved to higher embedding sizes (h = 64 to 256) and interaction layers (L = 3 to 5) as presented in **Table 1**. While standard deep learning conventions assume that deeper architectures yield higher fidelity, our results unveil a clear structural accuracy plateau. For the advanced tensor-equivariant models, scaling the network capacity beyond a critical threshold delivered marginal accuracy gains, and in specific regimes, actively deteriorated the force predictions. MACE and Allegro achieved near-peak performance at minimal hidden dimensions and shallow layers, most likely due to their handling of higher-body order interactions. MACE indeed explicitly constructs descriptive many-body atomic embeddings from scratch by expanding local atomic density into a spherical harmonics basis within a single layer. Allegro similarly bypasses global message passing altogether, updating environments through localized, layered tensor-product expansions. Consequently, both frameworks sidestep the structural necessity for deep network scaling.

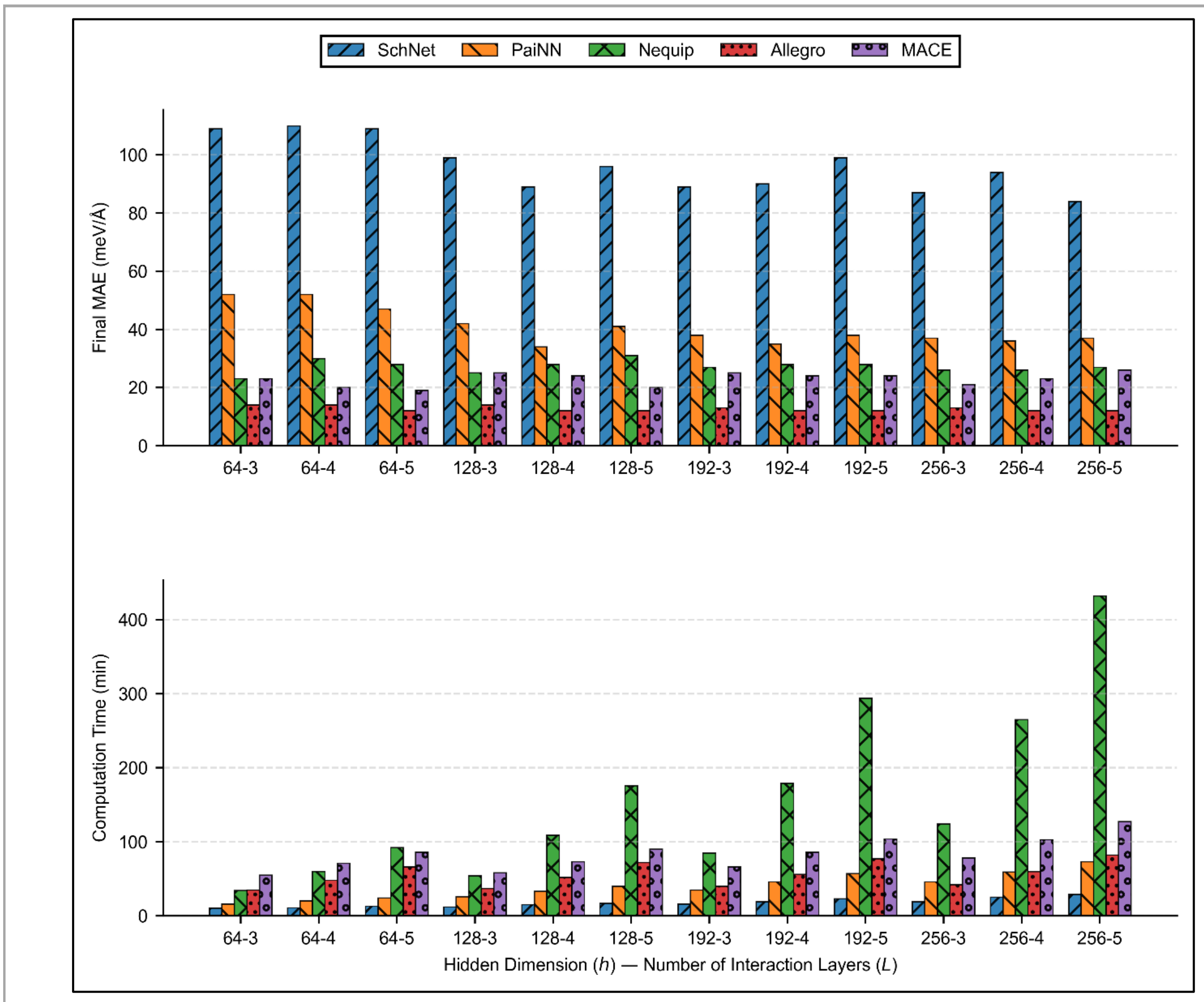

**Figure 2.** Accuracy and computational efficiency trade-offs for five GNN architectures (SchNet, PaiNN, Allegro, Nequip, and MACE) trained on $Cd_{68}Se_{55}Cl_{26}$ QD. The top panel illustrates the final force MAE (meV/Å), highlighting the predictive accuracy of each model. The bottom panel demonstrates the computational cost (minutes) required for training. Models are compared across varying network capacities, denoted by their hidden dimension (*h*) and number of interaction layers (L).

Conversely, traditional message-passing architectures exposed noticeable vulnerability to the typical pitfalls of GNN depth and capacity scaling. For instance, when increasing layers from L=3 to L=5 at h=128, the force MAE of NequIP degraded from 25 to 31 meV Å$^{-1}$. This performance drop is consistent with the effects of oversquashing, an informational bottleneck where aggregating expanding neighborhood graphs into fixed-dimensional vectors leads to a loss of localized structural information.

**Table 1.** Benchmarking results for $Cd_{68}Se_{55}Cl_{26}$ QD, showing final force MAE and computational efficiency across GNN architecture.

| Hidden Dimension | No. of Layers | SchNet | | PaiNN | | Nequip | | Allegro | | MACE | |
|---|---|---|---|---|---|---|---|---|---|---|---|
| | | Final MAE[a] | Time[b] | Final MAE | Time | Final MAE | Time | Final MAE | Time | Final MAE | Time |
| 64 | 3 | 109 | 10 | 52 | 16 | 23 | 34 | 14 | 35 | 23 | 55 |
| | 4 | 110 | 11 | 52 | 20 | 30 | 60 | 14 | 48 | 20 | 71 |
| | 5 | 109 | 13 | 47 | 24 | 28 | 92 | 12 | 66 | 19 | 86 |
| 128 | 3 | 99 | 12 | 42 | 26 | 25 | 54 | 14 | 37 | 25 | 58 |
| | 4 | 89 | 15 | 34 | 33 | 28 | 109 | 12 | 52 | 24 | 73 |
| | 5 | 96 | 17 | 41 | 40 | 31 | 176 | 12 | 72 | 20 | 90 |
| 192 | 3 | 89 | 16 | 38 | 35 | 27 | 85 | 13 | 40 | 25 | 66 |
| | 4 | 90 | 19 | 35 | 46 | 28 | 179 | 12 | 56 | 24 | 86 |
| | 5 | 99 | 23 | 38 | 57 | 28 | 294 | 12 | 77 | 24 | 104 |
| 256 | 3 | 87 | 19 | 37 | 46 | 26 | 124 | 13 | 42 | 21 | 78 |
| | 4 | 94 | 25 | 36 | 59 | 26 | 265 | 12 | 60 | 23 | 103 |
| | 5 | 84 | 29 | 37 | 73 | 27 | 432 | 12 | 82 | 26 | 127 |

[a] Force MAE [meV/Å]; [b] Computation time [min].

Concurrently, increasing the hidden dimensionality to h=256 led to localized performance degradation in MACE (26 meV Å$^{-1}$ at L=5, against the 20 at h=128), suggesting that increasing model capacity did not improve generalization for this dataset and may have increased sensitivity to the particular training configurations. Only Allegro demonstrated rigorous resilience to depth-induced oversquashing, holding its accuracy at 12 meV Å$^{-1}$ up to L=5.

As expected, scaling up the architectures heavily penalized computational throughput. The largest configurations took roughly two to five times longer to train than the smallest for most architectures, and more than 10-fold for NequIP, which scaled from 34 to 432 minutes, yet this additional cost brought no appreciable gain in accuracy. Consequently, we conclude that an optimized baseline configuration of a hidden dimensionality of 128 paired with four interaction layers offers the best balance between accuracy and computational load. Operating within this optimized regime effectively circumvents the prohibitive costs of over-parameterization and depth-induced informational losses while retaining the strict physical fidelity required to probe the complex potential energy surface of colloidal quantum dots.

### 3.2 Computational Cost and Inference Speed.

While data efficiency and robustness are critical, the practical application of these models relies heavily on their inference speed. Consequently, **Table 2** summarizes the inference time per frame for each architecture, benchmarked on the Orchestr.AI platform using 32 CPU cores and a single NVIDIA A100 GPU. A clear hierarchy in computational throughput was observed. SchNet exhibited the fastest inference, requiring 6 ms per molecular-dynamics step, but failed to maintain a physically valid trajectory. PaiNN and Allegro required 11 and 12 ms per step, respectively, but remained dynamically unstable within the tested active-learning budget. NequIP completed the full 1 ns trajectory without dataset augmentation and required 18 ms per step, making it the fastest model that was intrinsically stable under the tested conditions. Following the addition of 34 configurations selected through active learning, MACE also completed the 1 ns trajectory, although at a higher computational cost of 66 ms per step.

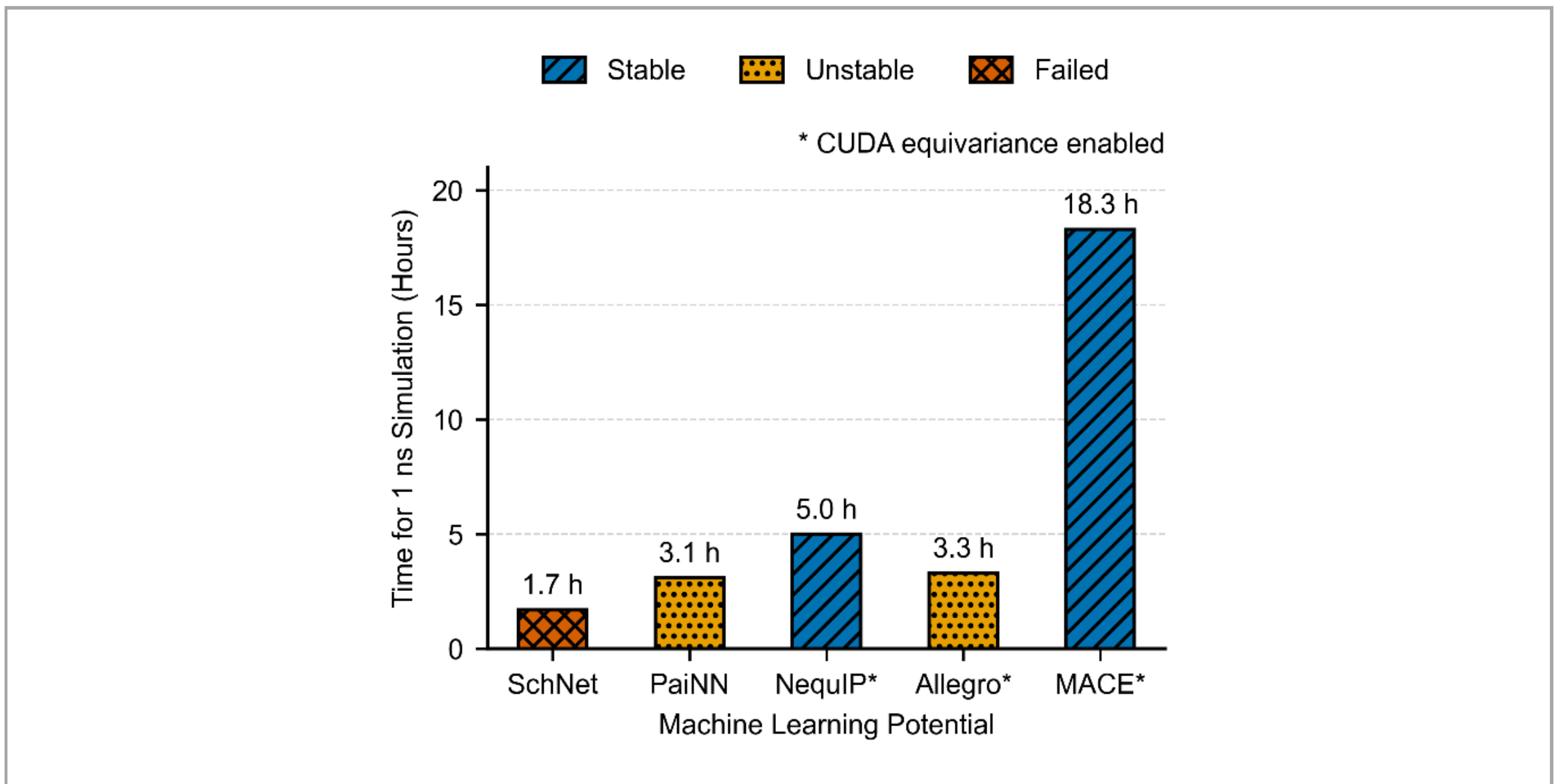

**Figure 3.** Performance trade-off of the machine-learned potentials relative to DFT: projected simulation time for a 1 ns trajectory (1.0 fs timestep) versus computational speedup relative to DFT for the five architectures. NequIP, Allegro and MACE were run with CUDA-accelerated equivariance

By comparison, a single AIMD step for the CdSe-Cl nanocluster requires approximately 20,000 ms (~20 s) on two nodes with 64 CPU cores and eight NVIDIA A100 GPUs, whereas each machine-learned potential was benchmarked on 32 CPU cores and a single A100 GPU. The machine-learning potentials evaluated here are therefore two to three orders of magnitude faster than the reference DFT calculation in wall-clock terms, and correspondingly more so on a per-GPU basis. Because the DFT and ML calculations used different numbers of GPUs and different hardware configurations, these values should be interpreted as wall-clock comparisons rather than strict per-GPU speedups. This reduction, however, in computational overhead bridges the gap between quantum-mechanical accuracy and the nanosecond-scale sampling required to resolve the structural dynamics of the nanocluster, as illustrated in **Figure 3**. These stability classifications should be interpreted within the tested protocol: one initial structure, 300 K, a Langevin thermostat, a 1 fs timestep and a 1 ns trajectory. They do not by themselves establish stability across different temperatures, structures or nanocrystal compositions.

**Table 2.** Computational efficiency and nanosecond-scale dynamical stability of the machine-learned potentials compared with DFT, for the $Cd_{68}Se_{55}Cl_{26}$ nanocluster (149 atoms). Timings are averaged over MD steps. The DFT reference was run on two nodes with 64 CPU cores and eight NVIDIA A100 GPUs, each machine-learned potential on 32 CPU cores and a single A100 GPU; NequIP, Allegro and MACE used CUDA-accelerated equivariance.

| Method | GPUs | Time per frame (ms) | Speedup | Stability |
|---|---|---|---|---|
| DFT (CP2K) | 8 | ~20,000 | 1× | Reference |
| SchNet | 1 | 6 | ~ 3,300× | Failed (40 ps) |
| PaiNN | 1 | 11 | ~ 1,800× | Unstable |
| NequIP | 1 | 18 | ~ 1,100× | Stable (1 ns) |
| Allegro | 1 | 12 | ~ 1,700× | Unstable |
| MACE | 1 | 66 | ~ 300× | Stable (1 ns) |

### 3.3 Dynamical Stability and Active Learning Refinement

To systematically evaluate and iteratively refine the potentials, the active learning workflow described in Section 2.3 was applied to sample out-of-equilibrium configurations across the CdSe potential energy surface. From the initial ensemble of 10 independent models trained for each architecture backend, the single model exhibiting the lowest force MAE was selected to drive an exploratory MD trajectory extended to a 1.0 ns timescale. For each saved frame, total energies and atomic forces were recomputed across all 10 ensemble members, allowing both the ensemble average and the epistemic uncertainty to be evaluated for every configuration. Among the frameworks evaluated, NequIP demonstrated the highest degree of inherent structural stability, completing the full 1.0 ns trajectory while maintaining a consistently low and stable uncertainty profile without requiring active dataset augmentation (**Figure 4g-i**). The remaining backends were subjected to targeted active-learning interventions to test whether additional configurations could improve their dynamical robustness (**Figure 5**). Allegro, despite the addition of 95 DFT-validated configurations, failed to reach a converged and stable trajectory, indicating that the uncertainty landscape of this potential could not be resolved through further data augmentation within the tested budget (**Figure 4j-l**). For PaiNN, the addition of 100 AL-sampled configurations reduced the frequency of high-uncertainty excursions but did not yield a converged, stable trajectory (**Figure 4d-f**). MACE was stabilized through the automated AL workflow after incorporating only 34 configurations into the training set (**Figure 4m-o**). In comparison, SchNet failed to maintain dynamical stability even over brief 40 ps durations, and no measurable improvement in robustness was observed after repeated AL cycles (**Figure 4a-c**).

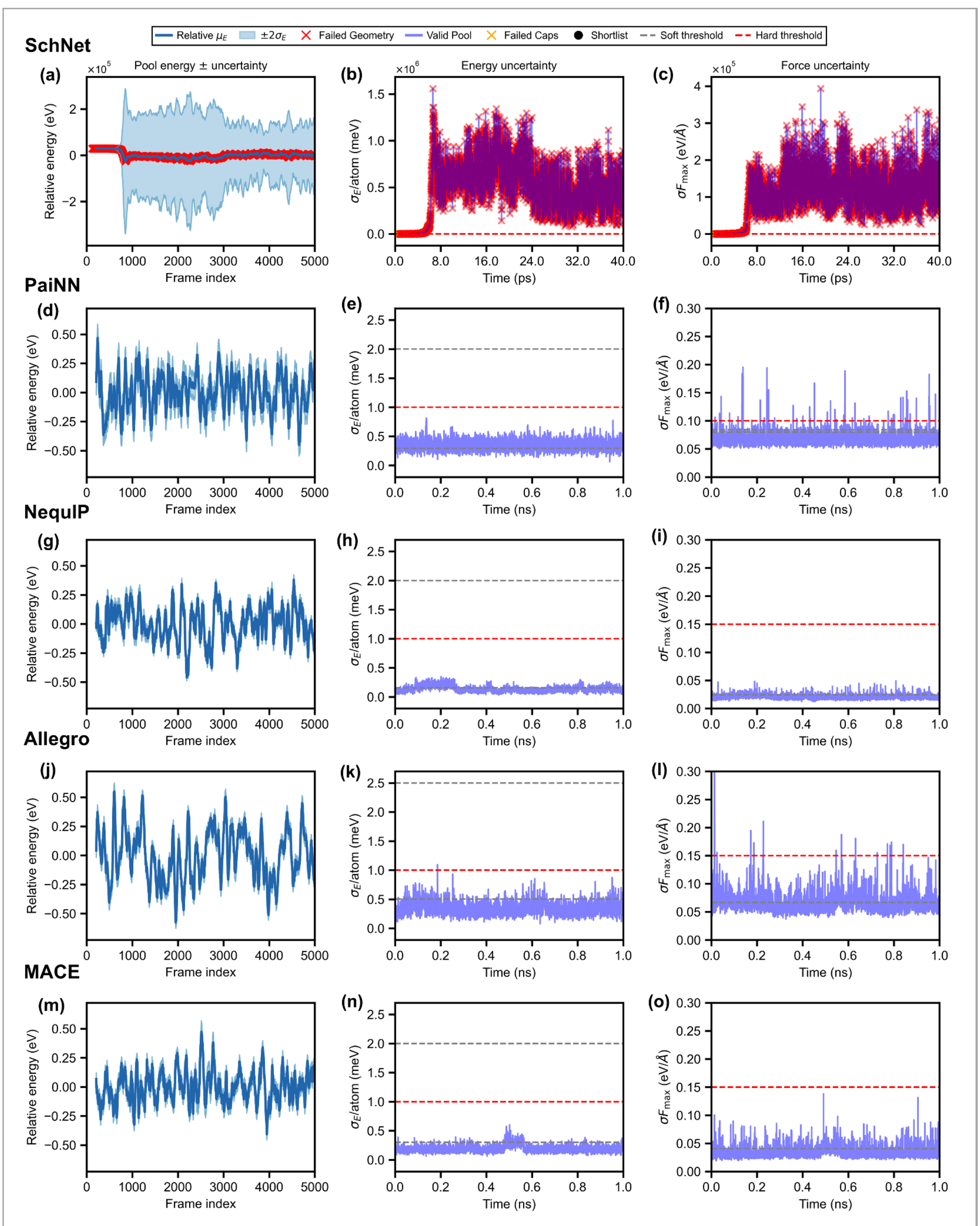

**Figure 4.** Active-learning diagnostics for five neural-network interatomic potentials: (a–c) SchNet, (d–f) PaiNN, (g–i) NequIP, (j–l) Allegro and (m–o) MACE. Columns show mean-subtracted pool energies with ±2σ_E uncertainty bands, per-atom energy uncertainties (σ_E/atom) with invalid frames marked separately, and maximum force uncertainties (σ_F,max) relative to soft (grey) and hard (red) thresholds. SchNet becomes unstable after approximately 5 ps and is plotted separately because its uncertainties are three to four orders of magnitude larger. PaiNN and Allegro repeatedly exceed their hard thresholds, whereas NequIP and MACE remain below them throughout the 1 ns trajectory.

Taken together, the results show that active learning can be highly effective for some architectures, particularly MACE, but that uncertainty-guided data augmentation does not guarantee dynamical stabilization within a fixed acquisition budget.

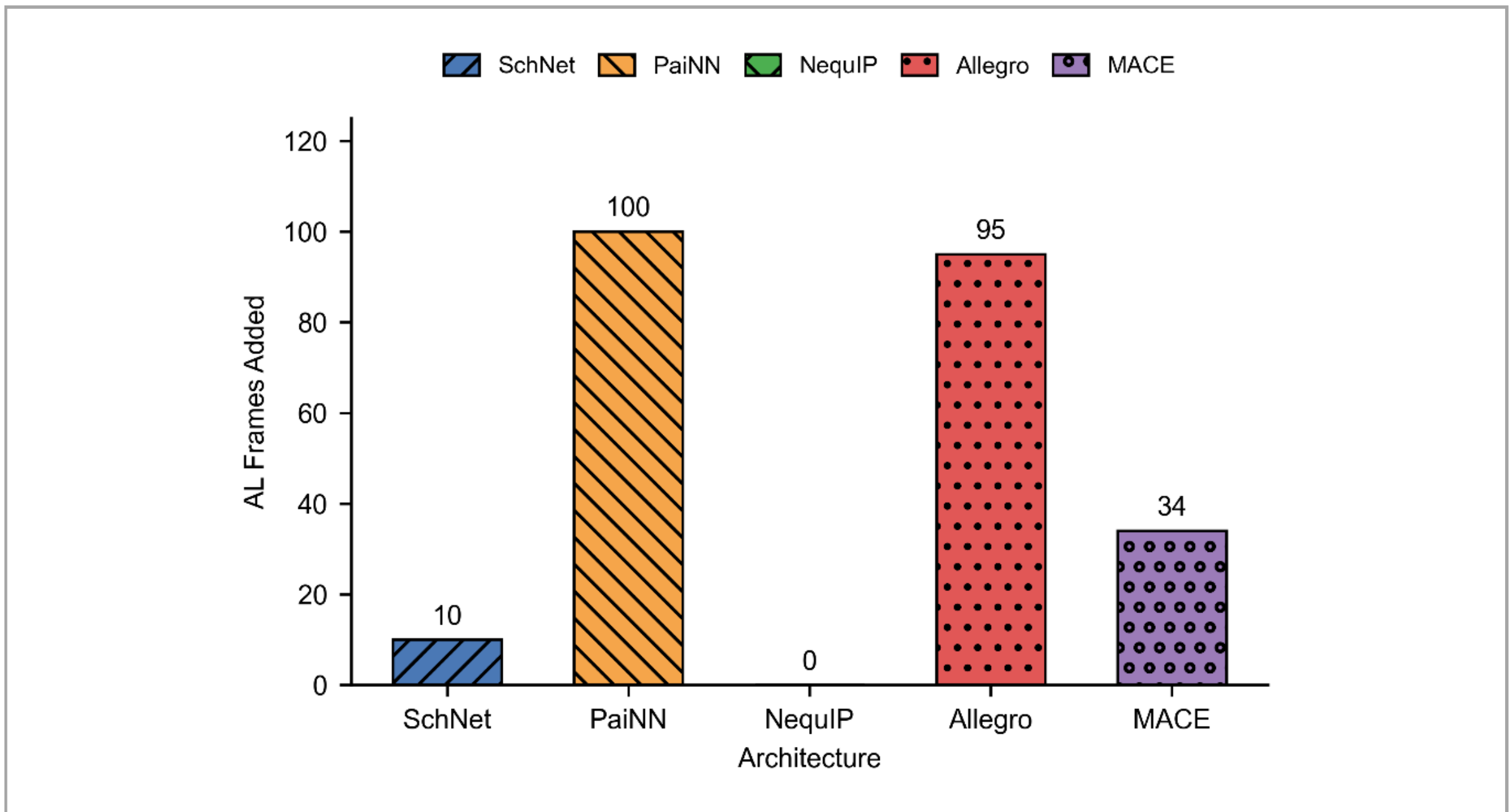

**Figure 5.** Number of DFT-labelled configurations added during the active-learning process for each neural-network architecture. A value of zero indicates that no augmentation was required.

## 4. Conclusions

This work establishes a harmonized framework for training and validating machine-learning interatomic potentials for a finite, surface-dominated semiconductor nanocluster. Benchmarking five graph-neural-network architectures for chloride-passivated CdSe demonstrates that low prediction errors on held-out configurations do not necessarily translate into stable long-timescale molecular dynamics. Allegro achieved the lowest force errors but remained dynamically unstable within the tested active-learning budget, whereas NequIP completed a 1 ns trajectory without additional training data. MACE was stabilized by adding only 34 uncertainty-selected configurations, demonstrating that active learning can efficiently improve dynamical robustness, although its effectiveness depends strongly on the underlying architecture. These results show that model selection should combine static accuracy, computational efficiency, uncertainty assessment and explicit dynamical validation. The conclusions are specific to the nanocluster and simulation conditions considered here, but the benchmarking protocol provides a reproducible basis for evaluating machine-learning potentials for other finite semiconductor nanostructures.

Because the present study considers a single 149-atom CdSe nanocluster with a defined morphology and chloride passivation, it should be regarded as a controlled starting benchmark rather than as the development of a universally transferable CdSe potential. Extending the models across a broader range of CdSe quantum dots will require training datasets that represent different sizes, shapes, crystallographic facets, surface stoichiometries and passivation patterns. This expansion can be performed iteratively by combining representative nanocrystal structures with active learning to identify local atomic environments that are insufficiently represented in the existing dataset. For larger nanocrystals, the training data could also include physically plausible surface reconstructions and dissociation or reattachment processes involving CdSe and $CdCl_2$ units. Such configurations would expose the models to undercoordinated atoms, distorted bonding environments and bond-breaking events that may not occur during short equilibrium trajectories.

Importantly, increasing the structural diversity of the dataset may not substantially reduce the average prediction errors, which are already low for several architectures under the present validation protocol. Its principal benefit may instead be to improve the physical consistency of sparsely sampled regions of the potential-energy surface and thereby enhance long-timescale dynamical stability. Data acquisition should therefore prioritize configurational diversity, uncertainty and configurations associated with trajectory failure rather than simply increasing the number of equilibrium structures. Fine-tuning pretrained foundation potentials offers a complementary route, provided that the underlying models contain an adequate representation of Cd–Se–Cl chemistry and are subsequently adapted using nanocrystal-specific surface configurations. Future work should compare such fine-tuning strategies with system-specific training and assess transferability across nanocrystal sizes, morphologies, temperatures and surface compositions.

## 5. Acknowledgments

We acknowledge Horizon Europe EIC Pathfinder program through project 101098649, named UNICORN by the European Union and research funding from the European Commission (MSCA-DN Track The Twin, grant agreement 101168820). We also acknowledge IKUR Strategy under the collaboration agreement between Ikerbasque Foundation and BCMaterials on behalf of the Department of Education of the Basque Government. The authors acknowledge the technical and human support provided by the DIPC Supercomputing Center. Computational resources were provided by the Hyperion cluster at the Donostia International Physics Center (DIPC). We acknowledge the EuroHPC Joint Undertaking for awarding this project access to the EuroHPC supercomputer LEONARDO, hosted by CINECA (Italy) and the LEONARDO consortium through an EuroHPC Access call.

## 5. Data Availability Statement

All data supporting the findings of this study are available from the corresponding author upon reasonable request. The training scripts and the automated Orchestr.AI framework used in this work are openly accessible at [https://github.com/nlesc-nano/Orchestr.AI](https://github.com/nlesc-nano/Orchestr.AI). Also, dataset for training and validation along with the ML models used in the benchmark are available at: [https://github.com/nlesc-nano/CdSe_Benchmark](https://github.com/nlesc-nano/CdSe_Benchmark)

## 6. Conflict of Interest

The authors declare no competing financial or personal interests that could have influenced the work reported in this paper.